\def\VEL{\:{\rm km\:s^{-1}}}

\documentclass[10pt]{article}
\usepackage{emulateapj}

\begin{document}


\newcommand{\MSOL}{\mbox{$\:M_{\sun}$}}


\title{{\it Far Ultraviolet Spectroscopic Explorer} Observations of the Supernova 
Remnant N49 in the Large Magellanic Cloud\altaffilmark{1}}
\author{William P. Blair\altaffilmark{1}, Ravi Sankrit\altaffilmark{1}, 
Robin Shelton\altaffilmark{1}, Kenneth R. Sembach\altaffilmark{1}, 
H. Warren Moos\altaffilmark{1}, John C. Raymond\altaffilmark{2}, 
Donald G. York\altaffilmark{3}, Paul D. Feldman\altaffilmark{1}, Pierre 
Chayer\altaffilmark{1}, Edward M. Murphy\altaffilmark{1}, David J. Sahnow\altaffilmark{1},
and Erik Wilkinson\altaffilmark{4}
}

\altaffiltext{1}{Department of Physics and Astronomy, The Johns Hopkins
University, 3400 N. Charles Street,  Baltimore, MD 21218}

\altaffiltext{2}{Harvard-Smithsonian Center for Astrophysics, 60 Garden St.,
Cambridge, MA 02138}

\altaffiltext{3}{University of Chicago, 5640 S. Ellis Ave., Chicago, IL 60637}

\altaffiltext{4}{University of Colorado, Campus Box 593, 1255 38th St.,
Boulder, CO 80302}


\begin{abstract}

We report a {\it Far Ultraviolet Spectroscopic Explorer} satellite
observation of the supernova remnant N49 in the 
Large Magellanic Cloud, covering the 905 -- 1187 \AA\ spectral region. A 
30\arcsec\ square aperture was used, resulting in a velocity resolution 
of $\sim$100 $\rm km ~ s^{-1}$.  The purpose of the observation was 
to examine several bright emission lines expected 
from earlier work and to demonstrate 
diffuse source sensitivity by searching for faint lines never seen 
previously in extragalactic supernova remnant UV spectra.  Both 
goals were accomplished.
Strong emission lines of O~VI $\lambda\lambda$1031.9, 1037.6 and
C~III $\lambda$977.0 were seen, Doppler broadened to $\rm \pm 225 ~ 
km ~ s^{-1}$ and with centroids red-shifted to 350 $\rm km ~ s^{-1}$, 
consistent with the LMC.  Superimposed on the emission lines are 
absorptions by C~III and O~VI 
$\lambda$1031.9 at +260 $\rm km ~ s^{-1}$, which are attributed to warm 
and hot gas (respectively) in the LMC.  The O~VI $\lambda$1037.6 line 
is more severely affected by overlying interstellar and $\rm H_2$ 
absorption from both the LMC and our galaxy.  N~III $\lambda$989.8 is not
seen, but models indicate overlying absorption severely attenuates this line. 
A number of faint lines from hot gas have also been detected,
many of which have never been seen in an extragalactic supernova 
remnant spectrum.

\end{abstract}

Subject headings: ISM: individual (N49) --- Magellanic Clouds --- 
shock waves --- supernova remnants --- ultraviolet: ISM

\section{Introduction}

Optical/UV spectra of supernova remnants (SNRs) contain emission lines 
from a broad range of ionization and excitation states of many 
intermediate mass elements. Observations with the {\it Voyager} Ultraviolet
Spectrometers (Blair et al. 1991a; Vancura et al. 1993, Blair et al. 1995a) and
the Hopkins Ultraviolet Telescope (HUT; Blair et al. 1991b; Long et al. 1992, 
Blair et al. 1995b) have revealed the stronger lines  expected in this 
spectral region.  However, the low spectral resolution of these earlier 
observations, contamination by residual airglow emission lines, and concerns 
about overlying absorption have compromised the utility of these data. 

A case in point is the bright X-ray and optical SNR known as N49 in the 
Large Magellanic Cloud (LMC; see Mathewson et al. 1983; Long et al. 1981).  
This object was observed with HUT (Vancura et al. 1992b), but 
the residual airglow in the sub-Ly$\alpha$ region
caused considerable difficulty in measuring the intrinsic SNR emissions.
Vancura et al. (1992b) were able to infer the presence and approximate total
strength of the O~VI  $\lambda\lambda$1031.9, 1037.6 doublet, but could not 
separate the lines clearly from the surrounding airglow.  

In this paper, we report {\it Far Ultraviolet Spectroscopic Explorer} 
satellite (FUSE) observations of N49.  These observations resolve
airglow emission from intrinsic SNR emission lines, and allow us to 
investigate intrinsic line profiles and overlying absorptions.  We also 
report a number of fainter emission lines expected from comparison with 
shock model calculations or coronal emissions from hot gas.  These 
observations demonstrate the capabilities of FUSE for line profile analysis
in SNRs and for detecting faint, diffuse emissions by virtue of its
sensitivity and extremely low background levels.

\section{Observations and Reductions}

The observations were obtained with the FUSE
satellite (Moos et al. 2000, Sahnow et al. 2000) during the in-orbit 
checkout period.  FUSE was pointed toward N49 on 1999 Sept. 21-22 UT, for 
34 ks total integration time, about 14 ks of which was during orbital night.  
We used the 30\arcsec\ square apertures to maximize throughput. 
However, the emission from many different filaments, with differing densities, effective
reddenings, and shock velocities, are being sampled simultaneously (cf. Vancura 
et al. 1992a,b).  Thus, the FUSE observation only samples the ``global" nature of 
the N49 emissions.

FUSE operates using four telescope channels which are intended to be co-aligned
for optimal performance.  At the early epoch of these observations, 
only three channels were approximately aligned.  
Figure 1 shows the location of the LiF apertures (solid line, aligned to 3\arcsec)
and the SiC2 aperture (dashed line) on an [O~III] optical image.
The SiC2 channel was offset such that the observed counts in the spectral 
overlap region are down by a factor of two from what would have been 
expected based on the relative effective areas of the channels. 
The SiC1 channel was not aligned, and contained only airglow emissions. 

The data reductions and calibrations are only approximate at this point,
consisting of either pre-flight or preliminary in-flight algorithms and
calibration values.  
We work primarily with the raw time-tagged ``counts" data sets, and 
collapse the lines into one-dimensional spectra from each of the three channels
that contained useful data.  We also applied a preliminary flux calibration to
assess approximate relative line intentisites. One can inspect subsets of these 
data (for instance, the ``night only" data) to eliminate transient events or to
identify features arising from airglow.
Because the resolution of the filled-aperture data is already degraded, we 
have simply co-added multiple channels whenever a region is 
covered by more than one.

\section{Discussion}

The FUSE spectrum of N49
demonstrates substantial advantages over previous observations in this spectral
region.  The SNR emission lines are not only separated from contaminating
airglow emission lines, but the kinematics and line profiles of the SNR lines
can also be assessed directly.  Also, the detection of a number of faint
emission lines that had not been detected previously in an extragalactic SNR
demonstrates FUSE's excellent sensitivity to faint diffuse emissions (cf. also 
Shelton et al. 2000).  This bodes well for future FUSE observations of nebular 
emissions that are not as intrinsically complicated as the situation we observe 
in N49.  Table 1 summarizes the
lines detected from the SNR and can be used to estimate the observed relative
line intensities.  Because some regions were observed by multiple channels and
the channels were not in perfect alignment, we list measured fluxes 
from separate channels for O~VI.  

\subsection{Strong Lines}

In Figure 2a, we show the spectral region from 1020 -- 1045 \AA\ from the
total and night only data sets, a region where all three active channels 
could be summed.  This region contains several airglow emission lines (from 
Ly$\beta$ and O~I) and the O~VI doublet from the SNR.  The airglow lines
fill the aperture, and indicate the filled slit spectral resolution,
which is 0.3 \AA\ or $\sim$100 $\VEL$.  
The SNR O~VI $\lambda$1031.9 line is separated from the airglow lines, while 
some minor contamination of O~VI $\lambda$1037.6 is caused by the O~I
$\lambda$1039.2 feature. (This contamination goes away in the night-only 
data section.) The SNR lines are much wider than the airglow features, 
broadened by the Doppler motion of material within the aperture.
The FWHM of the 1032 \AA\ line is 450 $\VEL$, consistent with
narrow slit optical echelle spectra (cf. Vancura et al. 1992a)
and a filled slit assumption.
Figure 2b shows the 970 -- 995 \AA\ spectral region for comparison.  (This
region was only sampled with a single SiC channel.)
The primary SNR line detected in this region is C~III 
$\lambda$977.0, although conspicuous by its absence is N~III $\lambda$989.8.
(This will be discussed below.)  

In Figure 3 we show these three strong lines on a velocity scale for comparison.
Several points are immediately obvious.  First of all, the lines are all 
considerably redshifted from rest velocity. 
With the airglow lines adjusted to their rest wavelengths, the centroids
of the O~VI lines are seen to be shifted by about +350 $\VEL$.
This is higher than the mean LMC velocity but consistent with previous observations
of N49 (cf. Shull 1983; Chu \& Kennicutt 1988).
Secondly, the observed profiles of the two O~VI lines, which should be intrinsically 
identical except for peak intensity, differ from one another.
This is because $\lambda$1031.9 lies in a spectral
region free of overlying $\rm H_2$ or interstellar absorption lines, while
$\lambda$1037.6 lies in a relatively complex region involving both $\rm H_{2}$ and
interstellar (especially C~II and C~II*) absorption, at both galactic and LMC velocities.
The $\lambda$1031.9 line, being less severely attenuated, is a better representation
of the intrinsic line profile .
Thirdly, comparing the C~III $\lambda$977.0 and O~VI $\lambda$1031.9 profiles,
the absorption dips align at $\sim$260 $\VEL$.
This is close to the expected mean redshift of the LMC, and we thus 
attribute these dips to C~III and O~VI absorption by diffuse LMC gas along
the line of sight to the SNR.  The absorption must be present in O~VI
$\lambda$1037.6 as well, but is largely masked by other overlying absorptions.
Any similar absorption from the Milky Way halo along this line of sight 
would be shortward of the blue wing of these lines.
It is clear that the C~III line is affected by overlying $H_{2}$ absorption,
especially on the red wing, where a plateau of emission is seen out to the
full extent of the O~VI velocity profile, but at a reduced level compared 
with O~VI.

The absence of N~III $\lambda$989.8 in the observed spectrum is
interesting.  This line traces the same ionization state and gas
temperature as C~III $\lambda$977.0.  Thus, in the
recombination zones behind radiative shocks, both lines are emitted
from the same region.  The relative intensity of the N~III line to the
C~III line is therefore a direct measure of the relative abundances of
N and C, and is less dependent on shock velocity or pre-shock density.

To obtain a measure of the N~III to C~III line ratio in N49, we ran
shock models using an updated version of the code described by
Raymond (1979).  These models were run with the LMC abundances
reported by Dufour, Shields, \& Talbot (1982), with C = 7.90 and N = 6.97 
(on a logarithmic scale with H = 12).  For shock velocities between 200 
$\VEL$ and 300 $\VEL$ and a pre-shock density of 30~cm$^{-3}$ (cf. Vancura et al. 1992a),
the N~III $\lambda$989.8 to C~III $\lambda$977.0 ratio is about 0.06 or less.
(Using abundances from Vancura et al. 1992a and a power law grid of models
yields an even lower expected ratio of $\sim$0.03.)


The region of the spectrum around 990~\AA\ has several H$_2$ lines
(Figure 2b) and interstellar lines that could absorb N~III line photons 
and make it effectively unobservable.  To investigate this possibility we ran
simulated FUSE spectra and then attenuated
them by assumed overlying absorption.  The C~III and N~III lines were
modeled as gaussians with FWHM $\sim$ 270 $\VEL$.  The peak C~III flux
was taken to be 
the observed value, and two cases for the N~III peak were considered
-- 0.1 and 0.5 times the C~III peak intensity.  We used two components 
for the absorbing gas, one at 0 $\VEL$ and the other at 250 $\VEL$,
corresponding to Milky Way and LMC velocities, respectively.  Each
component had a kinetic temperature of 250~K and an H$_2$ column of
10$^{18}$~cm$^{-2}$.  We find that this simple model for the absorbing 
gas severely attenuates the N~III line, completely absorbing N~III in
the weaker line assumption and nearly removing it in the stronger line assumption.

It is possible that other
interstellar lines contribute towards blanketing the N~III line.
In particular, based on Si~II lines at longer wavelengths
(cf. Welty et al. 1999), we expect Si~II $\lambda$990 to be a significant
absorber.  Also, self-absorption by LMC gas,
as seen in the other strong lines, should contribute to removing the
line.  

We note that similar attenuation must happen elsewhere in the spectrum
as well. The same model used above for N~III produces the attenuation
we see in the O~VI $\lambda$1037.6 line at least qualitatively. 
Using this model to make a first order correction to the strength
of the 1037.6 \AA\ line indicates that the intrinsic ratio of
O~VI $\lambda$1031.9/$\lambda$1037.6 is $\sim$1.7, somewhat lower
than the optically thin value of 2.0.  This indicates that some
of the O~VI emitting regions within the aperture are optically thick.

\subsection{Weak Lines}

One of the hallmarks of shock heated gas is the simultaneous presence of a broad
range of ionization states of many ions.  
The expected emission from the hotter components is similar
to that from coronal plasmas (cf. Feldman et al. 1997).

In Figure 4, we show several small spectral regions containing some of
the fainter lines from N49 detected with FUSE.  
[Ne~V] $\lambda\lambda$1137,1146, S~VI $\lambda\lambda$933,944, and
Ne~VI $\lambda\lambda$999,1006, are from temperature regions of 200,000 --
500,000 K (cf. Mazzotta et al. 1998), intermediate to X-ray emitting gas and 
regions sampled by the cooler ions observed at longer wavelengths, and similar to
O~VI.  However, the abundances of Ne and S are much lower than O, attesting
to the sensitivity of FUSE for observing faint, diffuse sources.  
Also shown in Figure 4 is a region near 1085 \AA, showing a faint line with 
possible absorption blueward from line center as in the stronger lines.
This is either the He~II $\lambda$1084.5, or the N~II $\lambda$1084.0 line
with self-absorption at the LMC velocity.
We also see evidence for a faint broad line near 1074 \AA\ which we identify 
with S~IV $\lambda$1073.0.  Another S~IV line expected
near 1063.5 \AA\ appears to be marginally present, but is expected to be attenuated
by overlying H$_2$ absorption.
Although many of these lines have been detected at a marginal level in 
HUT spectra, none of these faint lines have been 
detected previously in an extragalactic SNR.  With future detailed modeling,
these lines will provide new constraints on the cooling plasma behind
interstellar shock waves.

\acknowledgments

It is a pleasure to thank the hundreds of people at our collaborating
universities and institutions who worked on the development phase of FUSE,
and the many scientists and engineers who have checked out the instrument 
and are now operating it on orbit.  This work is based on data obtained 
for the Guaranteed Time Team by the NASA-CNES-CSA FUSE mission 
operated by the Johns Hopkins University. Financial support to U. S. 
participants has been provided by NASA contract NAS5-32985.


{\bf Figure Captions:}

{\bf Figure 1:}
Optical [O~III] CCD image of N49 in the LMC, from Vancura et al.
(1992a).  Overlaid as a solid box is the position of the
30\arcsec\ aperture viewed by the guidance camera on FUSE.
The offset dashed box shows the approximate position of the SiC2 channel, which is
more uncertain in the vertical direction than in the horizontal direction.
The SiC1 channel was not yet aligned at the time of this observation.
North is up and east to the left, and the spatial scale
can be derived from the aperture overlays.

{\bf Figure 2:}
Selected regions from the FUSE spectrum of N49 showing the
strongest lines detected. a) the region from 1020 -- 1045 \AA\ containing
the O~VI doublet, and b) the region from 970 -- 995 \AA\ covering C~III
$\lambda$977.0.  Note the absence of any obvious N~III $\lambda$991 emission.
Narrower lines arise from airglow emission lines and indicate the line widths
expected from a filled aperture.

{\bf Figure 3:}
The three strongest SNR lines shown on a velocity scale.  The middle
line corresponds to O~VI $\lambda$1031.9 and is least affected by overlying
absorption from $H_{2}$ and interstellar lines.  Shown on top is C~III $\lambda$977.0
and below is the other O~VI line.

{\bf Figure 4:}
Some of the fainter lines detected in the
FUSE spectra of N49.  a) S~VI $\lambda\lambda$933.4,944.5; b) Ne~VI
$\lambda\lambda$999.2,1005.7; c) [Ne~V] $\lambda\lambda$ 1136.5,1145.6, and
d) a blend of He~II and N~II near 1085 \AA.

\clearpage

\begin{deluxetable}{cccc}
\tablecaption{Observed line strengths 
                                   \label{TBL_SPEC}}
\tablewidth{0pt}
\tablehead{
  \colhead{Ion} & \colhead{$\lambda$} & \colhead{Flux} & \colhead{Channel}
}
\startdata
     C~III   &  977 &  $1.9\times10^{-13}$  & SiC 2A     \nl
      O~VI   & 1032 &  $5.8\times10^{-13}$  & Combined   \nl
      O~VI   & 1038 &  $2.2\times10^{-13}$  & Combined   \nl
\tablevspace{10pt}
      O~VI   & 1032 &  $6.4\times10^{-13}$  & LiF 1A   \nl
      O~VI   & 1038 &  $2.2\times10^{-13}$  & LiF 1A   \nl
      O~VI   & 1032 &  $5.7\times10^{-13}$  & LiF 2B   \nl
      O~VI   & 1038 &  $2.1\times10^{-13}$  & LiF 2B   \nl
      O~VI   & 1032 &  $5.5\times10^{-13}$  & SiC 2B   \nl
      O~VI   & 1038 &  $2.3\times10^{-13}$  & SiC 2B   \nl
\tablevspace{10pt}
      S~VI   &  933 &  $1.1\times10^{-14}$  & SiC 2A   \nl
      S~VI   &  945 &  $1.2\times10^{-14}$  & SiC 2A   \nl
     Ne~VI   & 1000 &  $1.5\times10^{-14}$  & LiF 1A   \nl
     Ne~VI   & 1006 &  $1.2\times10^{-14}$  & LiF 1A   \nl
He~II+N~II   & 1085 &  $2.7\times10^{-14}$  & SiC 2B   \nl
  {[Ne~V]}   & 1136 &  $0.9\times10^{-14}$  & LiF 1B   \nl
  {[Ne~V]}   & 1146 &  $1.6\times10^{-14}$  & LiF 1B   \nl
\enddata
\tablecomments{These are observed values, uncorrected for reddening
and absorption.  The units are erg~s$^{-1}$~cm$^{-2}$.}
\end{deluxetable}




\begin{references}
\begin{verse}

Blair, W. P., Long, K. S., Vancura, O., \& Holberg, J. B.  1991a, ApJ, 374, 202

Blair, W. P., et al. 1991b, ApJ, 379, L33


Blair, W. P., Vancura, O., \& Long, K. S. 1995a, AJ, 110, 312

Blair, W. P., Raymond, J. C., Long, K. S., \& Kriss, G. A. 1995b, ApJ, 454, L35

Chu, Y. -H., \& Kennicutt, R. C.,  1988, AJ, 95, 1111

Dufour, R. J., Shields, G. A., \& Talbot, Jr., R. J. 1982, ApJ, 252, 461

Feldman, U., Behring, W. E., Curdt, W., Schuhle, U., Wilhelm, K., 
Lemaire, P., \& Moran, T. M. 1997, ApJS, 113, 195


Long, K. S., et al. 1992, ApJ, 400, 214

Long, K. S., Helfand, D. J., \& Grabelsky, D. A. 1981, ApJ, 248, 925

Mathewson, D. S., Ford, V. L., Dopita, M. A., Tuohy, I. R., Long, K. S.,
\& Helfand, D. J. 1983, ApJS, 51, 345

Mazzotta, P., Mazzitelli, G., Coalfrancesco, S., \& Vittorio, N. 1998, 
A\&AS, 133, 403

Moos, H. W., et al. 2000, ApJ, (this volume)

Raymond, J. C. 1979, ApJS, 39, 1


Sahnow, D., et al. 2000, ApJ, (this volume)


Shelton, R., et al. 2000, ApJ, in preparation

Shull, P. O.  1983, ApJ, 275, 611

Vancura, O., Blair, W.P., Long, K. S., \& Raymond, J. C. 1992a, ApJ, 394, 158

Vancura, O., et al. 1992b, ApJ, 401, 220

Vancura, O., Blair, W.P., Raymond, J.C., \& Holberg, J.B. 1993, ApJ, 417, 663

Welty, D. E., Frisch, P. C., Sonneborn, G., \& York, D. G. 1999, ApJ, 512, 636

\end{verse}
\end{references}
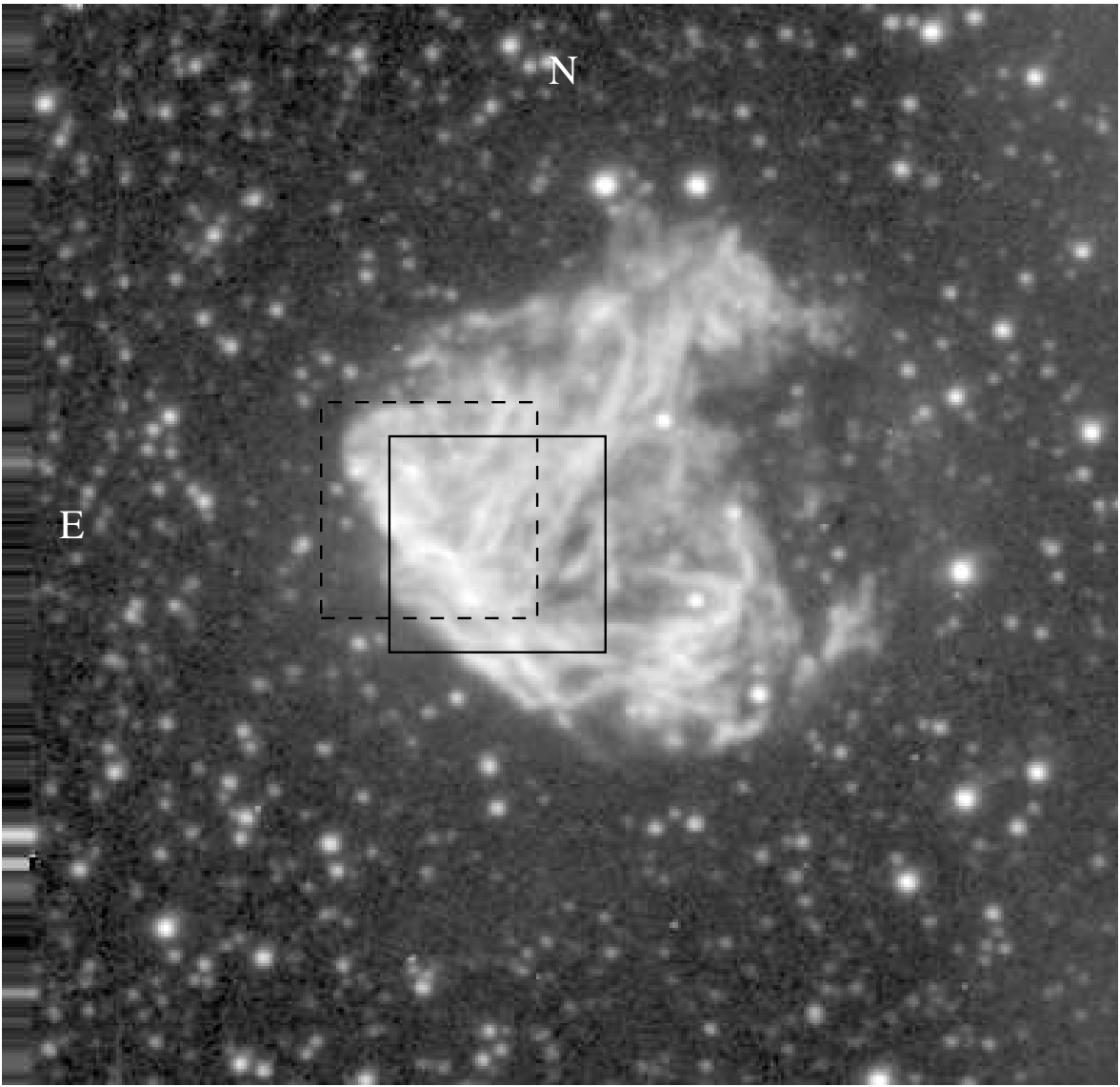
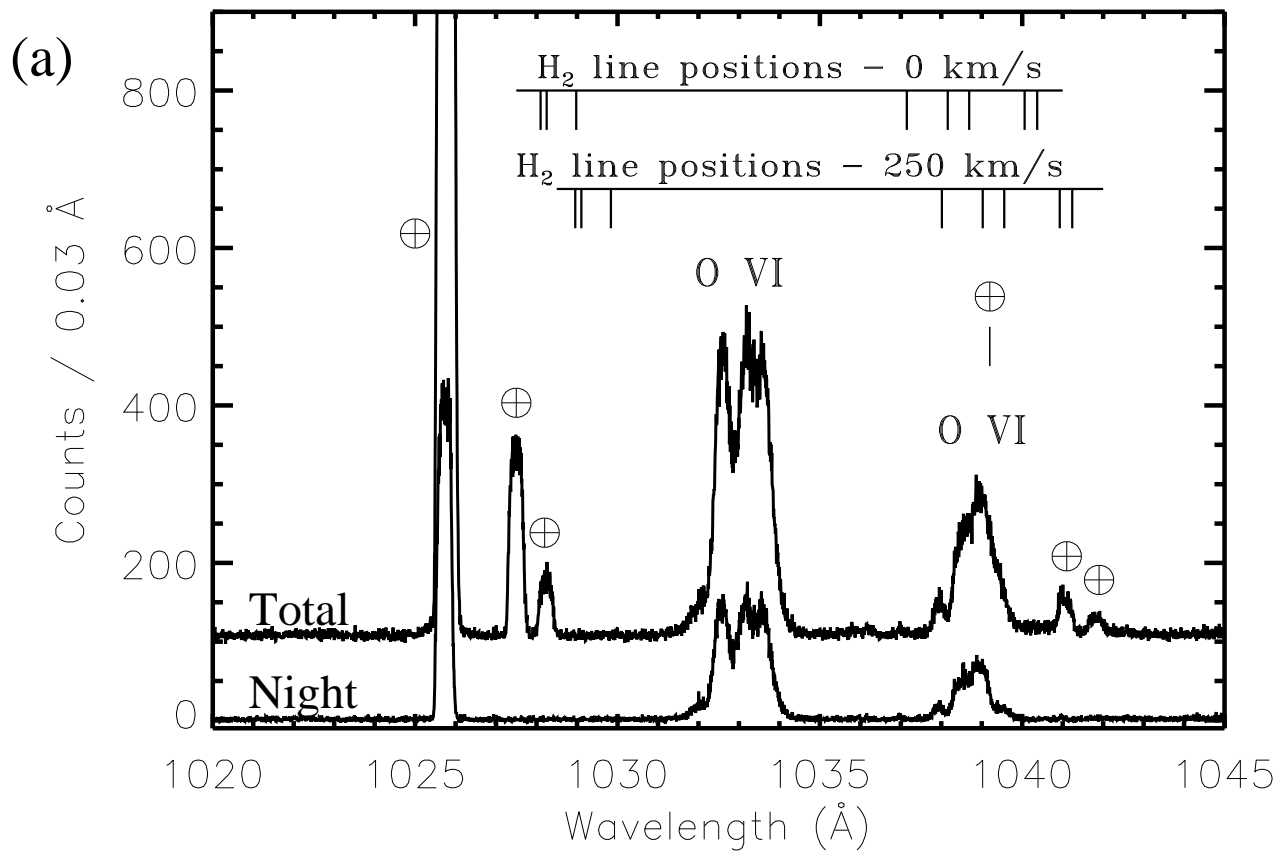
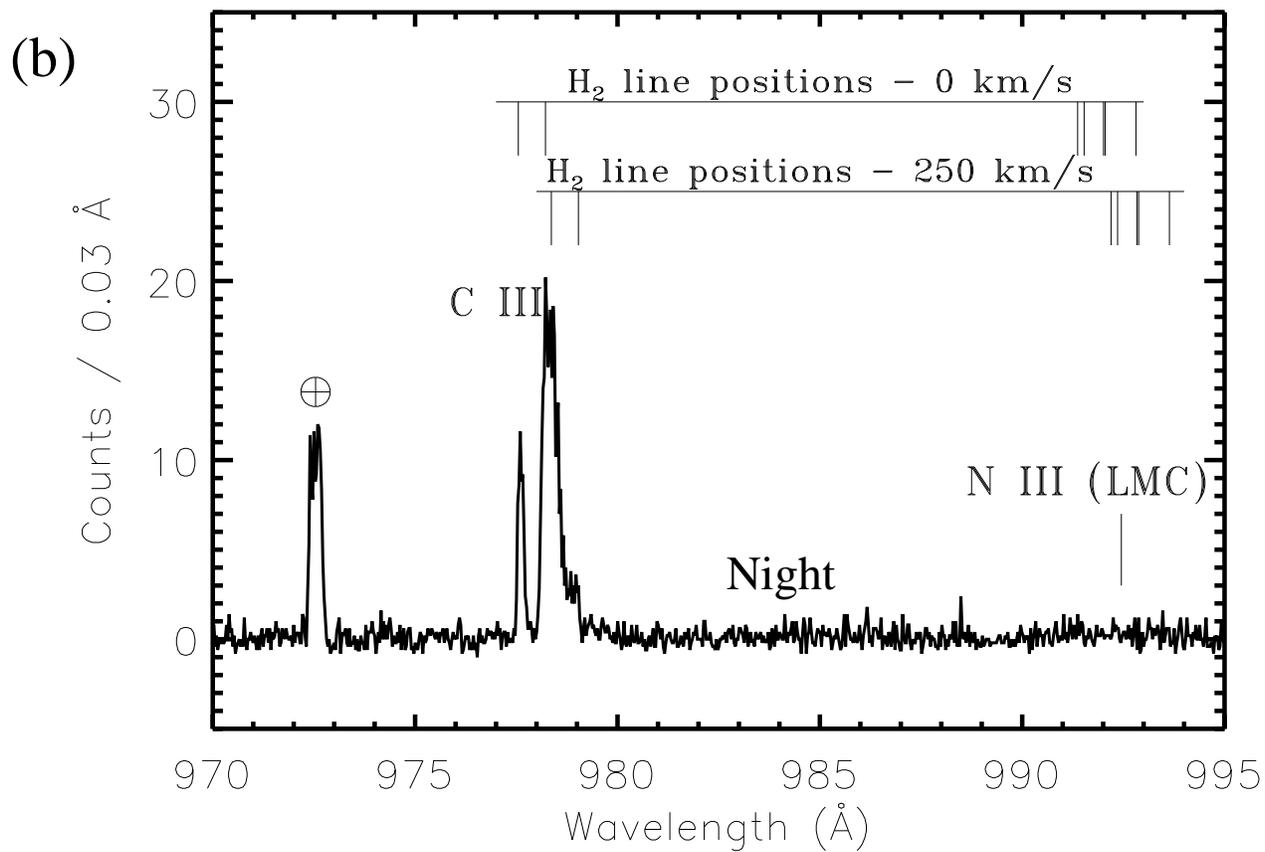
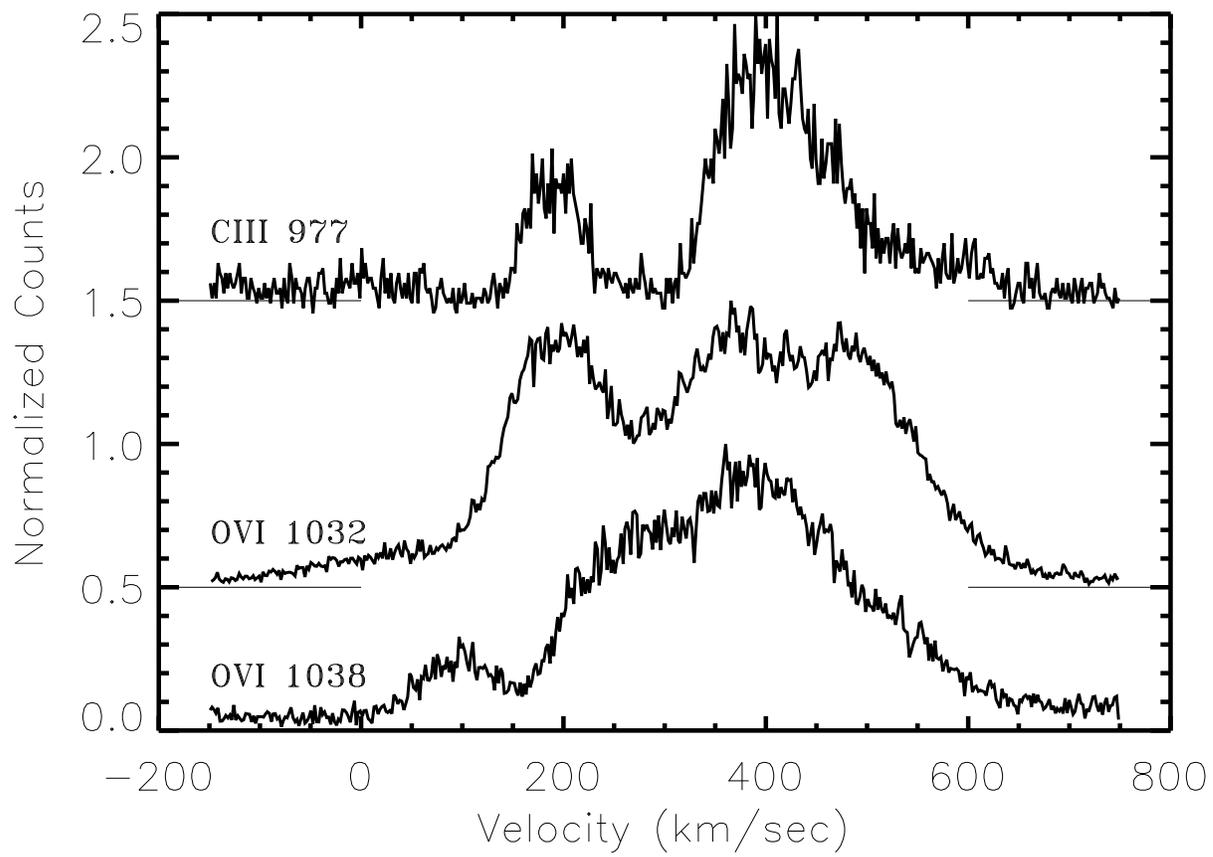
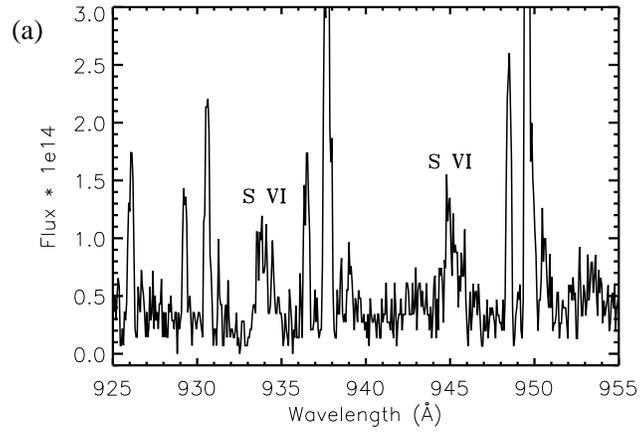
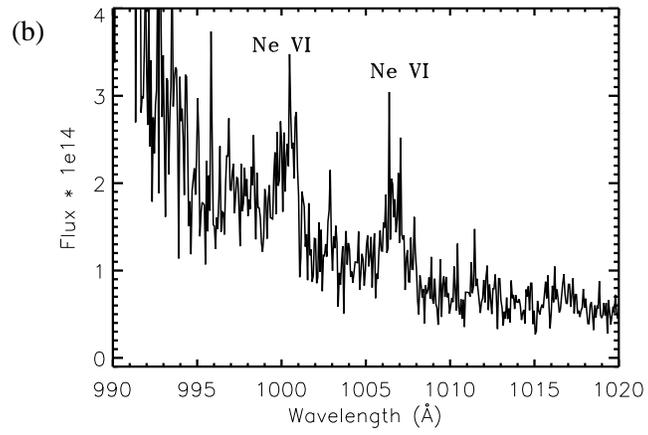
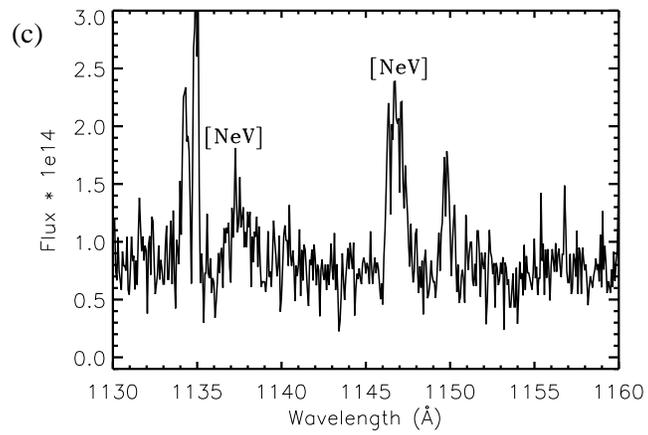
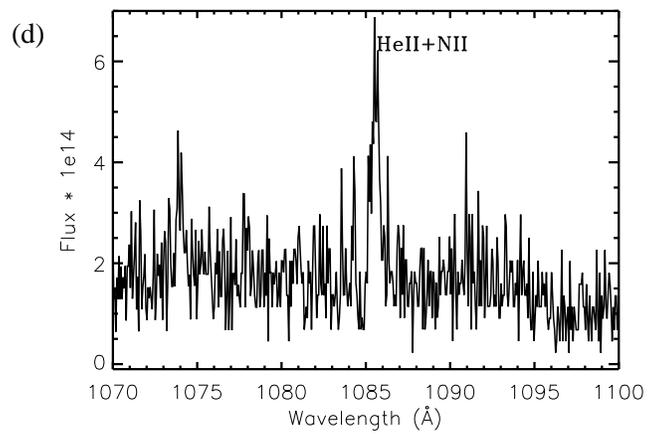
\end{document}